\documentclass[%
 reprint,
superscriptaddress,
nofootinbib,
 amsmath,amssymb,
 aps,
 showkeys,
]{revtex4-2}
\usepackage{graphicx}
\usepackage{dcolumn}
\usepackage{bm}
\usepackage{hyperref}
\usepackage[bottom]{footmisc}
\usepackage{footnote}
\usepackage{xcolor}
\usepackage[mathlines]{lineno}
\usepackage{fontenc}
\usepackage{graphicx}
\usepackage{dcolumn}
\usepackage{bm}
\usepackage{booktabs}
\usepackage{array,multirow}
\usepackage{amsmath}
\usepackage{amsfonts}
\usepackage{amssymb}
\usepackage{mathrsfs}
\usepackage{enumerate}
\usepackage{fancyhdr}
\usepackage{xcolor}
\usepackage{graphicx}
\usepackage{listings}
\usepackage{float}
\usepackage{verbatim}
\usepackage{braket}
\usepackage{centernot}
\usepackage{setspace}
\usepackage{endnotes}
\usepackage[normalem]{ulem}
\usepackage{wrapfig}
\usepackage{multirow}
\usepackage{comment}
\usepackage{kantlipsum}
\allowdisplaybreaks
\usepackage{lipsum, babel}
\hypersetup{breaklinks = true, colorlinks = true, citecolor = blue, linkcolor = blue, urlcolor = blue}
\usepackage[mathlines]{lineno}

\newcommand{\be}{\begin{equation}}
\newcommand{\ee}{\end{equation}}
\newcommand{\bea}{\begin{eqnarray}}
\newcommand{\eea}{\end{eqnarray}}

\newcommand{\LamZero}{$\Lambda~$}

\begin{document}

\title{Deep Exclusive Meson Production as a probe to the puzzle of $\Lambda$ hyperon polarization}

\author{Zhoudunming~Tu}
\email{zhoudunming@bnl.gov}
\affiliation{Department of Physics, Brookhaven National Laboratory, Upton, New York 11973, USA}

\date{\today}

\begin{abstract}
In the 1970s, an unexpected transverse $\Lambda$ polarization in unpolarized proton-Beryllium collisions was discovered, which initiated extensive studies on spin phenomena in high-energy physics. Over the past five decades, similar transverse $\Lambda$ polarization has been observed across various collision systems, including lepton-hadron deep inelastic scattering, hadron-hadron collisions, and electron-positron collisions. Despite numerous promising theoretical models, the fundamental mechanism underlying this polarization phenomenon remains inconclusive to this day. However, in both longitudinally and transversely polarized lepton-hadron and hadron-hadron collisions, it is found that the $\Lambda$ hyperon is \textit{not} polarized with respect to the initial parton spin direction. How the $\Lambda$ hyperon acquires its spin has become one of the most crucial questions to address in order to resolve this puzzle. In this paper, I propose to use an exclusive process that can be measured at the Electron-Ion Collider, the Deep Exclusive Meson Production, to explicitly test the mechanism of $\Lambda$ polarization. The outcomes of this experimental measurement are anticipated to unveil the dominant mechanism by which $\Lambda$ obtains its spin, eliminating many of the ambiguities that have been encountered in previous studies. Finally, experimental challenges and requirements will be discussed. 
\end{abstract}

\maketitle

\begin{figure*}
     \centering
     \includegraphics[width=.99\textwidth]{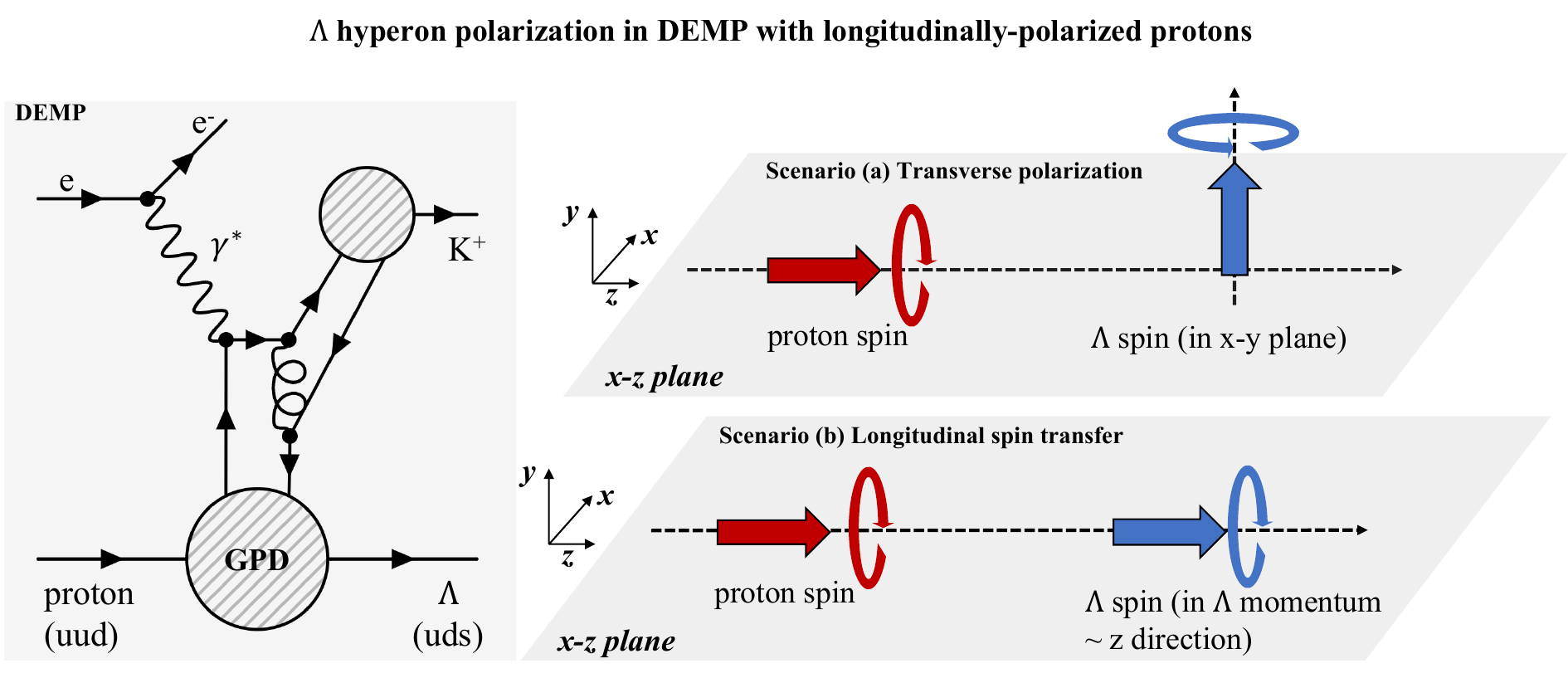}
     \caption{Illustration of Deep Exclusive Meson Production (DEMP) in a longitudinally polarized electron-proton scattering at the Electron-Ion Collider. The proton beam momentum can be between 41 to 275 $\rm{GeV/c}$, and the longitudinal polarization can be either positive or negative helicity. The \LamZero decay products do not need to be in the production plane, which is not shown.}
     \label{fig:kinematics}
 \end{figure*}

\section{Introduction}
Almost 50 years ago, a Fermilab experiment discovered a large transverse polarization of \LamZero hyperon in inclusive unpolarized proton-Beryllium collisions~\cite{Bunce:1976yb}. This was highly unexpected, because perturbative Quantum Chromodynamics (QCD) forbids a large polarization signal~\cite{PhysRevD.53.1073} and the total polarization in inclusive processes should average to zero in a naive expectation~\cite{Soffer:1991am}. This observation indicates that the \LamZero hyperon polarization has to originate from nonpertubative processes in a nontrivial way. Since then, transverse \LamZero polarization has been found in many different collision systems, e.g., electron-proton and electron-nucleus deep inelastic scattering (DIS)~\cite{HERA-B:2006rds,NOMAD:2000wdf,HERMES:2007fpi,ZEUS:2006ugq}, hadron-hadron and hadron-nucleus collisions~\cite{Bunce:1976yb,PhysRevD.40.3557}, heavy-ion collisions~\cite{STAR:2017ckg}, and recently even in electron-positron collisions~\cite{Belle:2018ttu}. 

Apart from the \LamZero polarization in heavy-ion collisions, which seems to be understood in the context of a strongly rotating Quark-Gluon Plasma~\cite{STAR:2017ckg,Liang:2004ph}, the origin of all other observed polarization remains inconclusive to-date. Many theoretical models~\cite{DeGrand:1980gc,DeGrand:1981pe,Joseph:1981zv,Soffer:1991am,Anselmino:2001js} have attempted to explain this phenomenon, and some were quite successful~\cite{DeGrand:1980gc,DeGrand:1981pe}, but there is no model that can explain the global data at the same time. Although all models point to a common direction, which is the hadronization process or final-state effects in general, the quantitative description of how \LamZero obtains its spin in unpolarized collisions is still unknown. 

What makes this problem even more interesting is when experiments attempted to measure \LamZero polarization in polarized target~\cite{COMPASS:2009nhs,STAR:2009hex,STAR:2018fqv,STAR:2018pps,Yu:2022tbj}. Thanks to the self-analyzing weak decay of \LamZero hyperon, \LamZero polarization measurement has been a common experimental tool to access the initial-state spin effect, e.g., the strange quark helicity and transeversity distributions. However, no \LamZero polarization has ever been found with respect to the initial spin direction of the target\footnote{Note that a nonzero polarization signal was found for $\bar{\Lambda}$~\cite{COMPASS:2009nhs}, which makes the whole picture more complex.}. It is expected that the \LamZero polarization asymmetry measurement is a convolution of parton distributions (e.g., helicity), parton scattering cross sections, and polarized fragmentation functions~\cite{Anselmino:2000ga}. Without understanding quantitatively the role of fragmentation, it is both experimentally and theoretically difficult to access the initial-state parton distributions. In other words, the hadronization process could be responsible for \textit{not} seeing the \LamZero polarization in polarized target~\cite{Anselmino:2000ga}. Recently, there are new proposals to measure \LamZero spin-spin correlation in both polarized and unpolarized deep inelastic scattering (DIS) and hadron-hadron collisions~\cite{Gong:2021bcp,Zhang:2023ugf,Vanek:2023oeo}, where final-state effects are expected to be suppressed. This is similar to the long-range two-particle momentum correlation in heavy-ion collisions~\cite{Li:2012hc}. There are other related studies based on \LamZero and its polarization at the EIC, e.g., leading \LamZero production~\cite{Carvalho:2023kfb} and inclusive transverse \LamZero polarization~\cite{Kang:2021kpt}.

In recent years, Deep Exclusive Meson Production (DEMP) has been proposed to be sensitive to Generalized Parton Distributions (GPDs) and meson form factors~\cite{osti_1362028,DEMPgen}. The process is as follows,  $e+p\rightarrow e'+K^{+}+\Lambda$, where the cross section of this process and its momentum transfer $-t$ are closely related to GPDs~\cite{osti_1362028}, thanks to the large $Q^{2}$ and high center-of-mass energy to ensure factorization in such a process. However, \LamZero polarization has never been considered in this context. Although this process was not initially intended for measuring the \LamZero polarization, it has a few advantages: i) an exclusive process with well-defined final states and kinematics; ii) a large \LamZero momentum that close to the beam direction, where the \LamZero unambiguously carries valence quarks from the incoming proton; iii) no feed-down from higher mass resonance and no fragmentation process involved. Therefore, this process would significantly simplify the picture of \LamZero polarization, which could potentially pin down its nonpertubative origin.

In this paper, a new \LamZero polarization measurement is proposed in the polarized $ep$ DEMP process at the upcoming Electron-Ion Collider (EIC). See Fig.~\ref{fig:kinematics} for an illustration. The measurement is designed to find out which direction the \LamZero hyperon is polarized: a) transverse polarization with respect to the production plane, or b) a large longitudinal spin transfer from proton to \LamZero, which is close to the direction of the incoming beams. Note that the spin direction of a) and b) are by definition orthogonal to each other. In this process, both scenarios are expected to have its possible maximum  strength. However, only one scenario can be correct. 

\section{Experimental techniques~\label{sec:technique}}

Experimental measurement of the \LamZero polarization is via its self-analyzing weak decay, $\Lambda \rightarrow p + \pi^{-}$. In the \LamZero rest frame, the momentum direction of the daughter proton (or $\pi^{-}$) exhibit a cosine distribution relative to the \LamZero polarization direction, as follows,

\be
\frac{dN}{d\cos{\theta}} \propto 1+\alpha P_{\Lambda}\cos{\theta}.
\ee

\noindent Here $\alpha$ is the weak decay constant~\cite{Ablikim:2018zay}, $\theta$ is the opening angle between the proton daughter in the \LamZero rest frame with respect to the \LamZero spin direction, and $P_{\Lambda}$ is the polarization magnitude of the \LamZero particle. The spin direction, however, can vary in different studies depending on the underlying mechanism of \LamZero polarization. In this study, I focus on two possible directions of \LamZero polarization: a) with respect to the production plane and b) with respect to the beam polarization (or \LamZero momentum) direction.

\begin{description}
    \item[Production Plane] The normal vector of the production plane is defined by $\vec{p}_{beam}\times \vec{p}_{\Lambda}$. In this study, the proton beam is used instead of the electron beam momentum as the $\vec{p}_{beam}$. 

    \item[Beam Polarization] For the longitudinal spin transfer, the expected \LamZero polarization is the direction of the \LamZero momentum. Some studies have tried to measure in the direction of the beam polarization, which is the same (or opposite) as the $\vec{p}_{beam}$ of the longitudinally polarized proton. In this study of DEMP, the direction of $\vec{p}_{\Lambda}$ and $\vec{p}_{beam}$ are very similar.
\end{description}

\noindent Therefore, it is by definition that the two spin polarization directions mentioned above are orthogonal to each other. Note that in the spin transfer measurement of polarized targets, it is common to use both helicity configurations to perform an asymmetry measurement, where most detector effects cancel. See Ref.~\cite{STAR:2018pps} for an example.  

\section{EIC experiments~\label{sec:eic}}
The upcoming EIC is designed to be an accelerator facility that can provide high energy and high luminosity electron-proton and electron-ion collisions, where proton and light ions can be also polarized~\cite{ref:EICCDR}. It will enable a comprehensive DIS measurements across a wide range of processes, including rare reactions that previously impossible to be measured at $ep$ collider experiments.

The current EIC project includes the accelerator facility and only one interaction region with one experiment - ePIC~\cite{ref:epic}. However, it is not excluded that there can be a second detector at the EIC. In fact, the EIC physics community favors two experiments~\cite{AbdulKhalek:2021gbh}. One potential difference of the second detector, among many other possibilities, can be the tagging capability in the hadron-going far forward (FF) region~\cite{FFwiki}.

Based on the current ePIC design, there are four detector subsystems in the FF region - the B0 spectrometer, Roman Pots (RP), Off-Momentum Detector (OMD), and Zero-Degree Calorimeter (ZDC). For details, see previous studies in Refs.~\cite{Tu:2020ymk,Chang:2021jnu,Jentsch:2021qdp}. Although the exact design has not been finalized yet, the general acceptance and performance are understood. Specifically, the B0 spectrometer can detector charged particles for scattering angles between 5.5 to 22.0 mrad. The RP detects the scattered proton in $ep$ collisions for small scattering angles, while the OMDs can detector breakup protons from nuclei with small scattering angles due to the change of the magnetic rigidity. Finally, the ZDC can detect neutral particles, e.g., the neutrons and maybe high-energy photons. 

For the process of interest in this study, the final-state particles are kaon and $\Lambda$, where the \LamZero decays to a pion and proton. The \LamZero will be close to the beam momentum with a small scattering angle. The challenge of detecting \LamZero from its decay is its long lifetime. From a current estimate~\cite{ref:EICCDR}, the acceptance of the \LamZero decay products in the FF system is small for high energy configuration, e.g., 18x275 GeV $ep$ collisions, but is feasible for 5x41 GeV (see a ePIC-like B0 simulation in Sec.~\ref{sec:simulation}). For \LamZero particles with an energy range of 30 to 41 GeV, the mean displaced vertex is at 2.7 meters downstream, which is before the location of the B0 spectrometer. The B0 is designed and expected to detect charged particle, e.g., protons and pions from the \LamZero decays. Quantitative estimation is needed and full detector simulations are being prepared in ePIC 
 and at the EIC second interaction point (IP). In addition, the kaon can be detected in the ePIC main detector with the barrel and the forward tracker, together with particle identification systems. An interesting direction for the EIC second detector is whether the acceptance of this measurement can be improved for high energy configuration, which allows access to the low-$x$ region. 

\section{Predictions}
\subsection{Transverse $\Lambda$ polarization with respect to the production plane}

From the measured data in the past decades, the transverse \LamZero polarization with respect to the production plane has been found with the following features: 
\begin{itemize}
    \item independent of the center-of-mass energy;
    \item strongly dependent on $x_F$ and $p_{T}$ of the \LamZero particle, and most data show a linear or quadratic dependence;
    \item the polarization is negative with respect to the production plane.
\end{itemize}

\noindent Note that the sign of the polarization signal depends on the definition of the production plane, which is usually $\vec{p}_{beam}\times \vec{p}_{\Lambda}$, where the $\vec{p}_{beam}$ is the momentum of the incident beam that carries valence up and down quarks. The polarization signal is found to have an opposite sign in lepton-nucleon (or nucleus) scattering (e.g., Ref.~\cite{HERMES:2007fpi}), and the natural explanation is that the incident beam is a lepton, which does not carry valence quarks. Therefore, the polarization signal is positive. It should also be noted that the transverse direction in the low-energy measurements within the resonance region~\cite{CLAS:2014udv} has a different definition of the projected spin axis. 

For hadron-hadron colliders, e.g., proton-proton collisions at RHIC and the LHC, both beams can be target and projectile, the choice is not unique. However, no polarization signal has been observed either~\cite{ATLAS:2014ona}.
Another feature to note is that the $\bar{\Lambda}$ polarization signal has never been observed in unpolarized hadron-hadron or lepton-hadron scatterings. 

In this proposed measurement, \LamZero particles are mostly in an extreme phase space. Specifically, the $x_F$ of \LamZero is close to -1, as the \LamZero will move forward close to the incoming proton beam momentum in the DEMP process. Note that the positive $x_F$ is defined as the longitudinal momentum fraction of a particle with respect to the incoming photon in the hadronic center-of-mass frame. However, to compare with previous proton-proton and proton-nucleus fixed target data~\cite{Bunce:1976yb,PhysRevD.40.3557}, the sign of $x_F$ is flipped in order to be compared properly. The naive picture is as follows: i) the proton has two up and one down quarks, together with sea quarks and gluons; ii) the strange and anti-strange  pair provides one strange quark to combine with one up and one down quark to form a \LamZero particle, while one valence up quark and the anti-strange quark form a positively charged kaon ($K^{+}$); both particles are moving forward, with \LamZero closer to the beam momentum. 

\begin{figure}
     \centering
     \includegraphics[width=.49\textwidth]{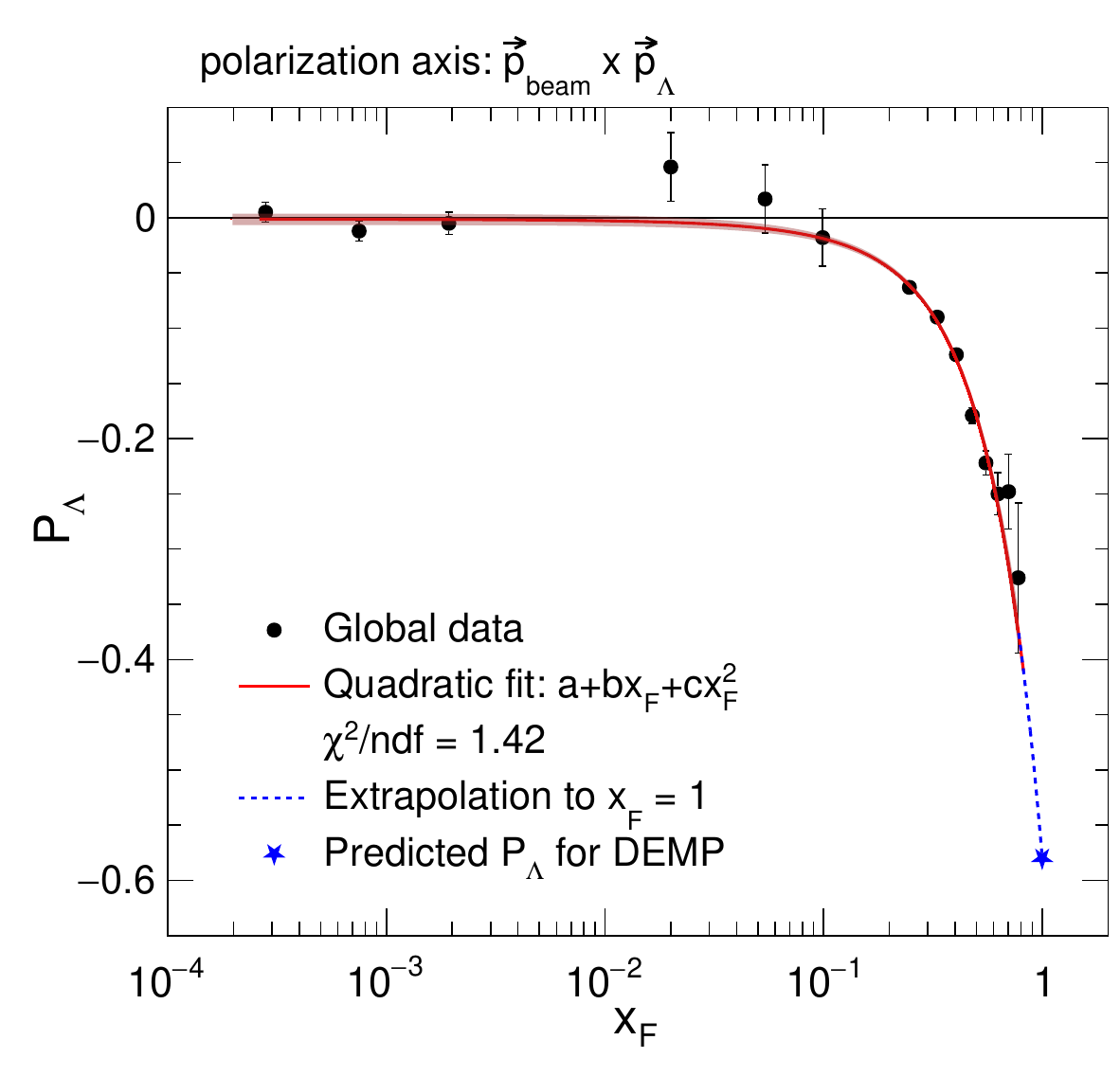}
     \caption{Transverse \LamZero polarization signal, $\vec{p}_{\Lambda}$, as a function of Feynman-$x$, $x_{F}$,  from the ATLAS~\cite{ATLAS:2014ona}, the HERA-B~\cite{HERA-B:2006rds}, and the M2 experiment~\cite{PhysRevD.40.3557}. A quadratic fit and the polarization signal extrapolated to $x_{\rm F}=1$ are shown. The sign of $x_F$ is flipped for DEMP process for proper comparison.}
     \label{fig:production-plane}
 \end{figure}

In Fig.~\ref{fig:production-plane},  data taken from ATLAS~\cite{ATLAS:2014ona}, HERA-B~\cite{HERA-B:2006rds}, and M2~\cite{PhysRevD.40.3557}, are shown as a function of $x_F$ of \LamZero particle. The data is fitted with a quadratic form, which describes the data really well with $\chi^{2}/ndf = 1.42$. Based on this extrapolation, the predicted $P_{\Lambda}$ value at $x_F=1$ is $-0.58$, which would suggest the dominant mechanism of \LamZero polarization is formed at the final state.  

Although the parametrization of a quadratic function is only to describe the data, the strong $x_F$ dependence is well established. There are many models that claimed to qualitatively describe this phenomenon, while some even claimed quantitative description. The one that has been most successful is the semi-classical quark recombination approach with the ``Thomas Precession"~\cite{DeGrand:1981pe,DeGrand:1980gc}. In this model, the polarization of the \LamZero particle is generated by accelerating the strange sea quark from low momentum ($x_{s}$) to roughly 1/3 of the \LamZero momentum. If the strange quark has a nonzero intrinsic transverse momentum $k_T$, the acceleration will naturally generate spin, e.g., 
\be
\omega_{\mathrm T}= \textbf{a} \times\textbf{v}_{\rm T}.
\ee

\noindent Here $\omega_{\mathrm T}$ is the Thomas Precession frequency, $\textbf{a}$ is the acceleration of the strange quark, and $\textbf{v}_{\rm T}$ is the transverse velocity of the strange quark. For high energy scatterings, the longitudinal momentum fraction of the strange quark can be very small, $x_s < 10^{-2}$, which corresponds to momenta less than 1 GeV. However, the \LamZero in DEMP can be produced up to hundreds of GeVs close to the beam momentum, where 1/3 of the beam momentum can be on the order of $\approx$ 100 GeV. Therefore, the expected acceleration of the strange quark is large at high energy, so are the $\omega_{\mathrm T}$ and polarization of the \LamZero particle.

There are other models that claimed to describe the transverse polarization of \LamZero in unpolarized lepton-lepton, lepton-hadron, and hadron-hadron collisions. In the DEMP process, it is expected that those based on high spin baryon resonances production~\cite{Joseph:1981zv}, single-pion exchange~\cite{Soffer:1991am}, and polarising fragmentation function~\cite{Gamberg:2021iat} should not play a role in this process. Based on all these information, the prediction is as follows:

\textit{\textbf{Prediction (a):} for the scenario (a), the \LamZero polarization would be as large as negative 60\% with respect to the production plane. }

\subsection{Longitudinal spin transfer via \LamZero polarization}

From polarized lepton-hadron and hadron-hadron collisions, the parton helicity distributions of quarks are well understood in the valence region. There are two general spin sum rules, known as ``Ji sum rule" and "Jaffe-Manohar sum rule", respectively, as follows:
\be
1/2 = \Delta\Sigma/2 + L_{q} + J_{g},
\ee
and
\be
1/2 = \Delta\Sigma/2 + \Delta G + l_{q} + l_{g}.
\ee
\noindent Despite the difference in their approach, there is no ambiguity in counting spin contributions from quarks. Other terms are related to angular and orbital angular momentum of quarks and gluons. 

In the process of DEMP, the production of \LamZero from incoming proton beam can be viewed as a result of knocking out a valence up quark and an addition of strange quark from the sea. In the valence region, the data has shown that the two up quarks account for 60\%, down quark 
accounts for $-30\%$, and strange quarks and anti-quarks account for very small fraction of the total nucleon spin~\cite{HERMES:2004zsh}.  Therefore, removing one up quark and adding one strange quark, naively would result in a 30\% reduction of the total spin. In other words, for longitudinal polarized proton target with 70\% initial polarization, 
\be
\mathrm {proton [uud] (70\%~polarized) \rightarrow \Lambda[uds] (\sim 50\%~polarized)}. 
\ee
To measure this polarization, the spin axis is the momentum direction of the \LamZero particle, which is close to the beam momentum and initial polarization direction (e.g., positive helicity configuration). There are models, e.g., Ref.~\cite{Ellis:1995fc}, that predict similar result of the longitudinal \LamZero polarization in the target fragmentation region; however, it is not in the exclusive channel that the $x_{\rm F}$ is much smaller.


Most of longitudinal and transverse spin transfer measurements using \LamZero particles were measured at small $x_F$ or mid-rapidity. In DEMP, the \LamZero momentum is close to the initial polarization direction, which is expected to maximize the spin transfer. In addition, similar to the case of polarization with respect to the production plane, the fragmentation process does not play a role here, and thus, a clearer picture of the spin transfer from nucleon to parton can be obtained. Based on these expectations, the prediction is as follows: 

\textit{\textbf{Prediction (b):} for the scenario (b), the \LamZero polarization would be as large as positive 50\% with respect to the \LamZero momentum or beam polarization direction. } 

\section{Standalone detector simulation \label{sec:simulation}}
In order to show the experimental feasibility of this proposed measurement, the standalone GEANT simulation~\cite{GEANT4:2002zbu} study is performed to evaluate the acceptance of \LamZero reconstruction in the planned B0 detector in the EIC far-foward region. Although this is not the exact B0 simulation in ePIC, this framework has been used in previous publications~\cite{Tu:2020ymk,Jentsch:2021qdp} before ePIC collaboration was formed and it is sufficient to show experimental feasibility. For details see Refs.~\cite{Tu:2020ymk,Jentsch:2021qdp}. 

Specifically, the B0 spectrometer consists of
silicon tracking planes (four layers) embedded in the first dipole magnet after the IP, which is about 5 meters downstream.
The angular acceptance of B0 is designed for reconstructing charged
particles with angles $5.5<\theta<20.0$ mrad, such as the proton and pion decayed from the \LamZero particle from DEMP. The GEANT simulation describes the \LamZero particle decays and their secondaries that propagate through the pre-defined detector volume of the B0 detector. The decay daughters, protons and pions in this study, would go through different layers of the silicon planes and leave signals, regarded as hits. By requiring at least 3 out of 4 hits, tracking reconstruction uses the EicRoot\footnote{https://github.com/eic/EicRoot} (which uses the GenFit package with Kalman Filter algorithm~\cite{Fruhwirth:1987fm}) to obtain the momentum of particles. In a single event, if two oppositely-charged tracks are reconstructed and combined to a \LamZero particle, this event is considered as accepted. 

\begin{figure*} [tb]   
  \includegraphics[width=.8\textwidth]{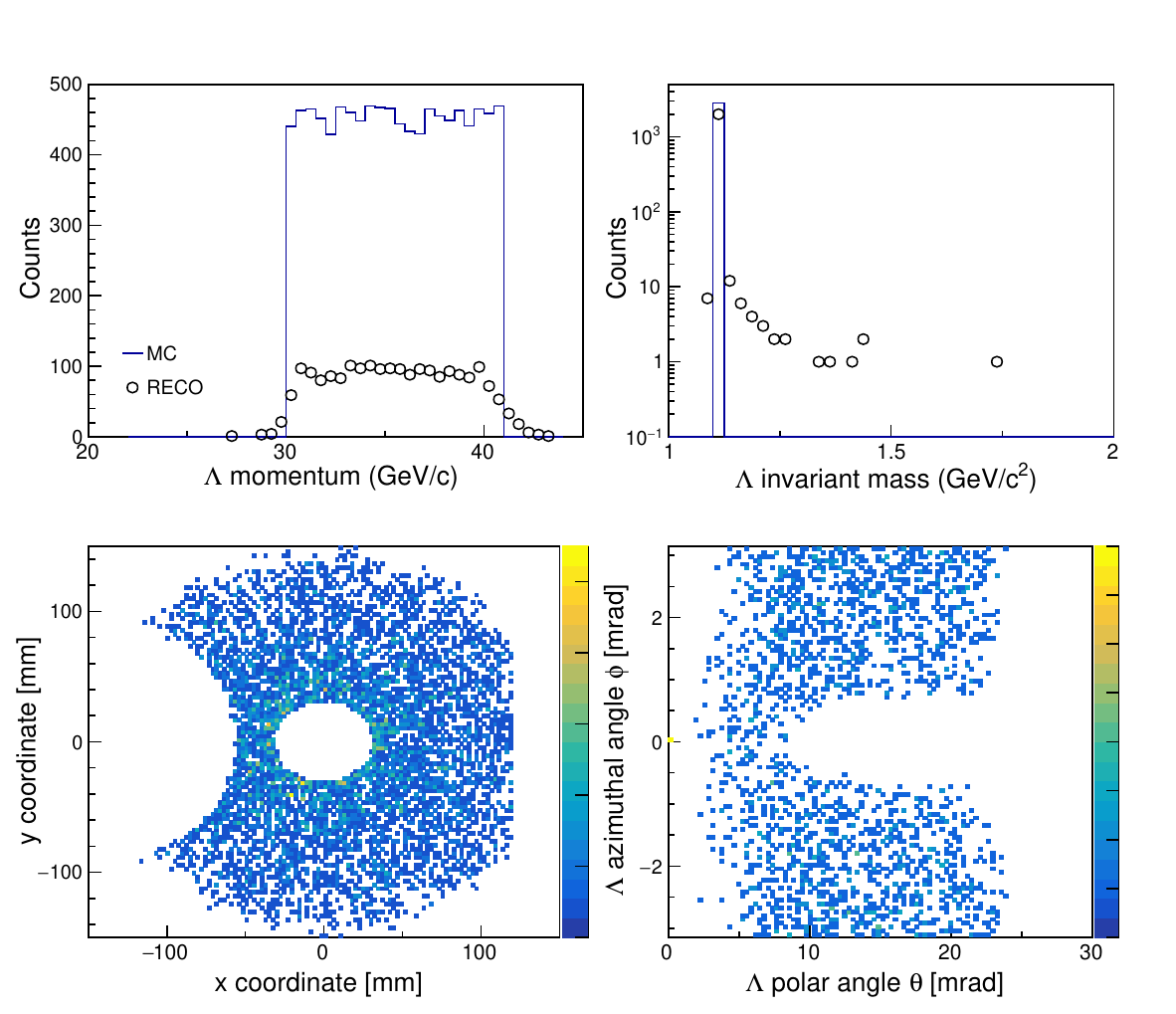}
     \caption{Standalone GEANT simulation of B0 for \LamZero reconstruction with a particle gun input between 30 to 41 GeV.}
     \label{fig:b0geant}
 \end{figure*}

In Fig.~\ref{fig:b0geant}, the \LamZero Monte Carlo (MC) truth and reconstructed momentum (top left), invariant mass (top right), hit signals detected in the B0 from \LamZero decays (bottom left), and the azimuthal verse polar angle of \LamZero (bottom right) are shown. The events are generated via a particle gun with flat distributions between 30 to 41 GeV/c in momentum as well as in scattering angle $\theta$ and azimuthal angle $\phi$. The reconstructed invariant mass distribution is peaked at the PDG \LamZero mass value but with a tail distribution at higher masses. This is caused mostly by the detector resolution. In addition, the reconstruction efficiency and acceptance are found to be $\approx 20\%$, largely due to the geometrical acceptance loss from electron and hadron beam pipes, quadruple magnet, and shielding. This can be seen from the bottom left panel of Fig.~\ref{fig:b0geant}, which shows the local coordinate of the B0 detector that the origin (0,0) is the center of the hadron beam pipe. The acceptance losses from the negative $x$ direction are due to the electron beam pipe, quadruple magnet, shielding, etc. These acceptance losses can be seen from the polar angle and azimuthal angle distributions, which are shown in the bottom right panel of Fig.~\ref{fig:b0geant}. Note that the 20\% acceptance includes contributions from all decay channels. Furthermore, most of the displaced vertices are within 5 meters from the IP, which is before the location of the B0 detector. From the same study with higher energy \LamZero particles, the acceptance is found to be significantly reduced because the positions of the displaced vertices are largely beyond the B0 detector. Nevertheless, the \LamZero reconstruction in the B0 at the low energy configuration, e.g., 5x41 GeV, is found to be feasible.  

\section{Discussion}
\begin{figure*} [tb]   
  \includegraphics[width=.8\textwidth]{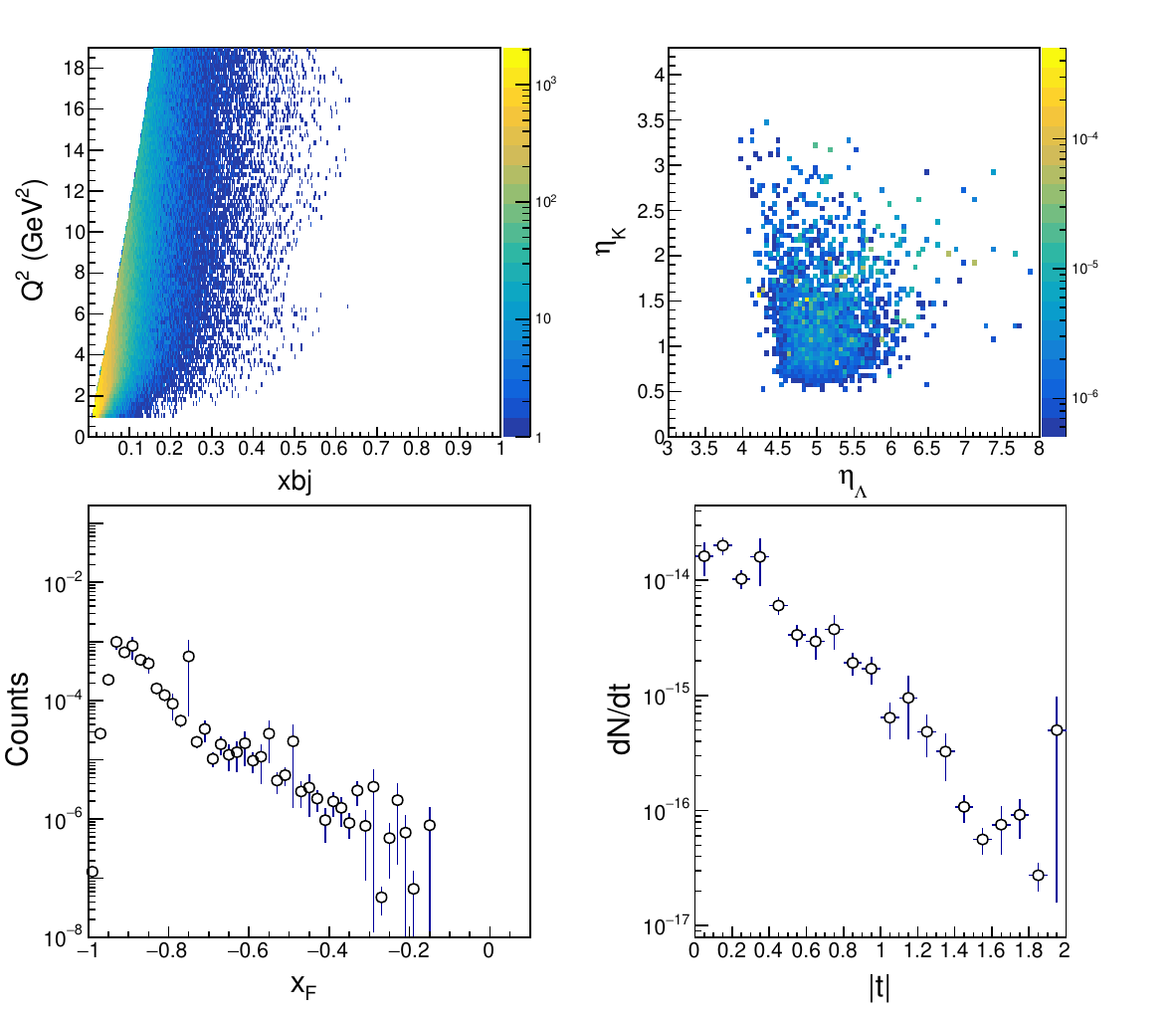}
     \caption{Kinematic distribution of Deep Exclusive Meson Production in the $e+p\rightarrow e'+K^{+}+\Lambda$ process, generated by the event generator DEMPgen~\cite{Ahmed:2024grm} }
     \label{fig:DEMPgen-kine}
 \end{figure*}

Based on the DEMP generator (Ref.~\cite{Ahmed:2024grm}), the rate of this process for 5x41 GeV at the EIC is around 18 Hz for the kaon-\LamZero final state. In other words, it is only about 10 $\rm pb^{-1}$ for generating $5\times10^{5}$ \LamZero for this study. See Fig.~\ref{fig:DEMPgen-kine} for the kinematic distributions of this process. The physics model is based on the exchange of Regge trajectories in the $t$ channel and constrained by the VGL model~\cite{Guidal:1997hy,Vanderhaeghen:1997ts,Guidal:1999qi,Guidal:2003qs} that describes exclusive hadronic reactions in the DIS region with low momentum transfer. The model has been compared with another model VR~\cite{Vrancx:2014pwa}, which was found to be reasonably close in describing the differential cross section in momentum transfer. Quantitative estimate of the model uncertainty is currently unavailable. 

The top-left panel shows the kinematic distribution $Q^{2}$ verse $x_{bj}$, where the large lever arm in $Q^{2}$ is one of the advantages of this process at the EIC. The top-right panel shows the pseudorapidity ($\eta$) distribution of kaon and the \LamZero particle in the lab frame. The forward going \LamZero is concentrated around 4--6 in $\eta$, which is within the acceptance of the EIC far-forward detector system in general (both ePIC and the conceptual design of a second detector). The bottom-left and bottom-right panel show the longitudinal momentum fraction with respect to the virtual photon momentum in the hadronic center-of-mass frame ($x_{\rm F}$) and the momentum transfer $-t$. The $x_{\rm F}$ distribution shows that the \LamZero is carrying most of the momentum of the proton beam going in the forward direction. 

Based on requirements of measuring \LamZero polarization in DEMP via its decay and the beam polarization, this measurement can be best, if not only, performed at the EIC. In terms of the luminosity requirement for 5x41 GeV of $ep$ collisions, an integrated luminosity of $\approx10~\rm{pb^{-1}}$ can result in a measurement with statistical precision better than 2\%, which is based on an estimate of the branching ratio of \LamZero decaying to proton-pion final state and the reconstruction efficiency and acceptance of \LamZero and kaon detection. This requirement is orders of magnitude lower than GPD measurements based on the same channel~\cite{Ahmed:2024grm}. 

Based on the GEANT simulation study above, it would be challenging to perform this measurement at higher energy configurations, e.g., 18x275 GeV $ep$ collisions, in ePIC given the location of the B0 detector. From the perspective of a EIC second detector, where the design of the detector and technology choice are not decided yet, a more downstream spectrometer away from the IP can benefit the higher energy \LamZero reconstruction in the far-forward region. In addition to the acceptance, the momentum and spatial resolution are expected to be good enough based on the current B0 design in ePIC to determine the direction of the \LamZero momentum reasonably well~\cite{Tu:2020ymk,Jentsch:2021qdp}. 
Quantitative feasibility studies at ePIC and the EIC second detector are needed and have been planned.

Taking one step further, this measurement can be extended to i) transverse polarized $ep$ DEMP and ii) the incoherent electron-deuteron or electron-helium 3 DEMP on the bound nucleon target. For i), the transverse spin direction is not always perpendicular to the production plane direction. One could vary the relative angle between them by selecting different azimuthal angle of the \LamZero particle in the forward direction. For ii) with Helium 3 polarization and the proton spectators tagging technique~\cite{Tu:2020ymk,Jentsch:2021qdp,Friscic:2021oti}, one could measure a similar process on the polarized neutron target to understand the role of valence quarks, e.g., $e+He^{3}(d)\rightarrow e'+2p'(p')+K^{0}+\Lambda$, where the neutral kaon decays weakly (e.g., $K^{0}_{s}$ to two pions). Experimentally, this is much more challenging than the proton case. In addition, one of the possible upgrades at the EIC is to polarize deuteron, where similar measurements can be done using the deuteron target.

Finally, experiments in Ref.~\cite{osti_1362028} and CLAS12 at Jefferson Lab with the 12 GeV electron beam program can measure the momentum transfer cross section of DEMP, based on a technique by utilizing the missing mass of the \LamZero particle. However, dedicated studies are needed to investigate the feasibility of measuring the \LamZero polarization in this process. Note that similar measurements in the resonance region with electron beam polarization have been done at CLAS before~\cite{CLAS:2014udv,CLAS:2022yzd}, while longitudinal polarized target data was not available until recently. A potential measurement with this data in CLAS12 is also interesting. At both Jefferson Lab 12 GeV program and at the EIC, this experimental measurement may be sensitive to Chiral-Odd GPDs~\cite{Beiyad:2010qg}, which may shine new lights to the problem of \LamZero polarization.

\section{Conclusions}
In conclusion, a new experimental measurement of \LamZero polarization in the process of Deep Exclusive Meson Production in polarized $ep$ collisions at the upcoming Electron-Ion Collider is proposed. This measurement is expected to probe directly the underlying mechanism of how \LamZero acquires its spin in high energy particle scattering processes. Specifically, this proposal has a number of advantages: i) a clean exclusive reaction with only kaon and \LamZero in the final state; ii) the \LamZero particle carries the \textit{maximal} longitudinal momentum, where maximum polarization with respect to both the production plane and the beam polarization direction are expected; however, only one can be correct; iii) no feed-down from higher mass resonances and no fragmentation involved in this process, which automatically rules out many theoretical models on transverse \LamZero polarization. The result of this measurement will significantly broaden the experimental program at the Electron-Ion Collider, e.g., using \LamZero hyperon as a probe to study Transverse-Momentum Dependent parton distribution functions and Generalized Parton Distributions. 
Most importantly, it may also provide a clear path forward towards solving the almost 50-year puzzle of \LamZero polarization. 
\\
\section*{Acknowledgements} 
The author would like to thank Alex~Jentsch for information and simulation for the Far-Forward detector system at the EIC. I would like to thank Garth~Huber and Wenliang Li for discussion on the Deep Exclusive Meson Production (DEMP) and the usage of DEMP event generator. I want to thank Francesco~Bossu, Xiaoxuan Chu, Abhay~Despande, Christian~Weiss, and the local BNL group for general discussion on this topic. The work is supported by the U.S. Department of Energy under Award DE-SC0012704 and the Laboratory Directed Research and Development (LDRD) 22-027 and LDRD-23-050 project.

\bibliography{demp-draft}

\end{document}